\documentclass[11pt]{article}
\usepackage{textcomp}
\usepackage{pdfsync}
\usepackage{fancyhdr}
\usepackage{amssymb}
\usepackage{amsmath}

\usepackage{graphicx}
\usepackage{latexsym}
\usepackage{appendix}
\usepackage{srcltx}
\def\CE{{\cal E}}

\def\CH{{\cal H}}

\def \w {\omega}

\def\centeron#1#2{{\setbox0=\hbox{#1}\setbox1=\hbox{#2}\ifdim
   \wd1>\wd0\kern.48\wd1\kern-.48\wd0\fi
   \copy0\kern-.48\wd0\kern-.48\wd1\copy1\ifdim\wd0>\wd1
   \kern.48\wd0\kern-.48\wd1\fi}}

\newcommand{\beq}{\begin{equation}}
\newcommand{\eeq}{\end{equation}}
\newcommand{\bea}{\begin{eqnarray}}
\newcommand{\eea}{\end{eqnarray}}
\newcommand{\ba}{\begin{array}}
\newcommand{\ea}{\end{array}}

\newcommand{\nn}{\nonumber}

\newcommand{\half}{\frac{1}{2}}

\begin{document}

\hskip3cm

 \hskip12cm{CQUeST-2011-0405}
\vskip3cm

\begin{center}
 \LARGE \bf  Quasi Normal Modes for New Type Black Holes \\
in  New Massive Gravity
\end{center}

\vskip2cm

\centerline{\Large Yongjoon
Kwon\footnote{emwave@khu.ac.kr}\,,~~Soonkeon
Nam\footnote{nam@khu.ac.kr}\,, ~~Jong-Dae
Park\footnote{jdpark@khu.ac.kr}\,, ~~Sang-Heon
Yi\footnote{shyi@sogang.ac.kr}}

\hskip2cm

\begin{quote}
Department of Physics and Research Institute of Basic Science, Kyung
Hee University, Seoul 130-701, Korea$^{1,2,3}$

Center for Quantum Spacetime, Sogang University, Seoul 121-741,
Korea$^4$
\end{quote}

\hskip2cm

\vskip2cm

\centerline{\bf Abstract} We obtain  the quasi-normal mode
frequencies of scalar perturbation on new type black holes in  three
dimensional new massive gravity. In some special cases, the exact
quasi-normal mode frequencies are obtained by solving scalar field
equations exactly.  On some parameter regions, the highly damped
quasi-normal mode frequencies are obtained in an analytic form by
the so-called Stokes line method.  This study on quasi-normal modes
sheds some light on the mysterious nature of these black holes.  We
also comment about AdS/CFT correspondence  and the entropy/area
spectrum for new type black holes. \thispagestyle{empty}
\renewcommand{\thefootnote}{\arabic{footnote}}
\setcounter{footnote}{0}
\newpage

\section{Introduction}

Three dimensional massive gravity model introduced by Bergshoeff,
Hohm and Townsend (BHT) leads to some renewed  interests in three
dimensional higher curvature gravity, since this so-called new
massive gravity (NMG) is composed of the standard Einstein-Hilbert
term with a specific combination of scalar curvature square term and
Ricci tensor square
one~\cite{Bergshoeff:2009hq}$\sim$\cite{Bergshoeff:2010mf}. One of
the original motivation of this theory is the non-linear completion
of Fierz-Pauli massive graviton theory or the realization of the
massive gravity on three dimensional space. Though NMG has different
nature, for instance parity even rather than odd,  from
topologically massive gravity
(TMG)~\cite{Deser:1982}\cite{Deser:1981wh}, these theories share
common features in some aspects.   One of the most interesting
common aspects in these theories is the existence of (warped) $AdS$
black hole
solutions~\cite{Bergshoeff:2009aq}\cite{Clement:2009gq}$\sim$\cite{Ghodsi:2010}.
Though the incompatibility of the unitarity of massive graviton
modes and $AdS$ black hole solutions is not resolved, it was
suggested that the special case may exist as a consistent quantum
gravity~\cite{Strominger:2008}.

Aside from massive gravitons, $AdS$ black hole solutions give us
another motivation to study NMG in the viewpoint of the $AdS/CFT$
correspondence, as was done extensively in
TMG~\cite{Nutku:1993}$\sim$\cite{Anninos:2008fx}. One direction
along the $AdS/CFT$ correspondence, {\it a la} holographic
c-theorem, is the extension of NMG to even higher curvature
gravities, named as extended NMG~\cite{Sinha:2010}\cite{Gullu:2010}.
In these cases, various $AdS$ black hole solutions are found and the
central charges of their conjectural dual CFT are
identified~\cite{Nam:2010dd}$\sim$\cite{Alishahiha:2010iq}. On the
contrary, there are different aspects in NMG from TMG, one of which
is the existence of the new type black holes discovered by BHT in
NMG which we will name {\it new type black holes} in the
following~\cite{Bergshoeff:2009aq}. These black holes exist only for
a specific combination of parameters in the NMG Lagrangian.

New type black holes are asymptotically $AdS$ and contain two
parameters even in the static case.  From the standpoint of the
usual Einstein-Hilbert gravity, the existence of an additional
parameter  is mysterious for  static and spherically symmetric black
holes (or axi-symmetric ones in the three dimensional case). Since
NMG is a higher derivative theory, the existence of an additional
parameter may be regarded as natural when one considers an initial
value problem. Though  the masses of these black holes are obtained
as a conserved charges  in the form of the combination of black hole
parameters, and the validity of the formula is  supported by the
first law of black hole thermodynamics and also by the dynamical new
type black hole solutions,  it is still unknown what is the physical
interpretation of additional parameter. Suppose that the additional
parameter or another combination of two parameters different from
mass combination is identified as a conserved charge, it is puzzling
that the first law of black hole thermodynamics  is already
satisfied without a new conserved charge. Moreover, mass seems to be
the unique conserved quantity for static new type black holes from
the symmetry consideration.

If the additional parameter is not a conserved charge, it is
desirable to study non-conserved quantities relevant to black hole
physics. One of these non-conserved dissipating quantities in black
hole physics is quasi-normal
mode(QNM)~\cite{Cardoso:2009}$\sim$\cite{Kokkotas:1999}(See
references there in). Some information about black holes can be
obtained by perturbing them and studying ensuing behaviors. Since
black holes are not closed systems, there are dissipating modes when
black holes are perturbed. The dissipating modes in black holes may
be defined by modes such as only ingoing part at the horizon and
only radiating one at the infinity. These modes with a specific
boundary condition are called QNMs. The same term is used for modes
of dissipating fields other than metric perturbation in the black
hole background. In this paper we study QNM frequencies, which are
complex numbers, of scalar fields on the new type black holes.

There are several suggestions how to define QNMs on black holes. One
of them is  to take   the boundary conditions such as ingoing modes
only at the horizon and outgoing modes only at the spatial infinity
with complex frequencies. In the asymptotic $AdS$ case, which is
relevant to us,   the Dirichlet boundary condition instead of
outgoing one may be taken  at the spatial infinity because of the
existence of blowing up potential~\cite{Horo:99}. QNMs mean the
decay of modes for the time elapse. Because of the complex nature of
QNM frequencies, ingoing/outgoing boundary conditions lead to
exponentially blowing behaviors of modes at the horizon and the
spatial infinity for fixed time. By separation of variables, the
radial part of field equations is given in the form of
\[  \bigg[ -\frac{d^2}{dx^2} - \omega^2  + U(x)\bigg] \Phi(x) =0\,.\]
In the viewpoint of the radial equations, ingoing/outgoing boundary
conditions mean taking the exponentially blowing part while ignoring
exponentially vanishing one. This process may be ambiguous since we
should take care of exponentially small parts which are usually
difficult to control in approximate solutions or asymptotic series
solutions of differential equations,  though one may not confront
such difficulties in exact solutions.

To overcome the ambiguity in boundary conditions taken in the domain
of real values, it was proposed to perform the analytic continuation
of radial coordinate to the complex domain and to impose boundary
conditions along Stokes lines~\cite{Schiappa:2004}. Stokes lines are
defined by the contour such that would-be vanishing or blowing terms
are equally contributing.\!\footnote{In mathematical literatures
these contours are called anti-Stokes lines.}  In our notation these
lines can be defined by ${\frak Im} (\omega x)=0$. On these lines
the boundary conditions can be imposed unambiguously. It may be
useful to note that one needs some additional information about
$\omega$ actually to draw  Stokes lines  on the complex plane of the
radial coordinate. Along with these boundary conditions,  a method
is developed  to obtain QNM frequencies using Stokes lines. This
method is basically monodromy approach to QNMs, while one needs some
modifications in the asymptotic $AdS$
case~\cite{motl:2003}$\sim$\cite{Ghosh:2005}.

In the asymptotic $AdS$ case Stokes lines are not closed and so open
lines are used to match approximate solutions.
This Stokes line method gives us approximate results since it uses
matchings among the approximate solutions through the interpolation.
However, some important outcome in QNM frequencies can be obtained.
For instance, the asymptotic QNM frequencies can be reliably
identified, which is important in the area spectrum and entropy
quantization through Hod's
conjecture~\cite{Hod:1998}\cite{magi:2007}.  In our case of new type
black holes, one can see some clues, through QNM frequencies, about
which combination of parameters different from mass is appropriate.

This paper is organized as follows. In section 2, we give a brief
review on the properties of new type black holes in NMG to fix our
conventions. In section 3, we solve the scalar field equations on
new type black hole backgrounds and obtain exact QNMs in some
special cases, which include results in the case of the
Ba\~{n}ados-Teitelboim-Zanelli (BTZ) black holes~\cite{btz}. In this
section we have taken the simple form of boundary conditions for
QNMs. In section 4, we obtain highly damped QNMs in some general
case using Stokes line method.  We conclude with summary and several
comments in the final section.   In the appendixes, we present some
mathematical formulae and numerical computation of QNM frequencies
for some parameter domains.


\section{New Type Black Holes in NMG}

\subsection{New massive gravity : brief review}

After Bergshoeff, Hohm and Townsend have introduced a higher
curvature theory (NMG) as the non-linear completion of Fierz-Pauli
massive graviton theory in three
dimensions~\cite{Bergshoeff:2009hq}$\sim$\cite{Bergshoeff:2009fj},
of which Lagrangian consists of the standard Einstein-Hilbert term
and a specific combination of scalar curvature square term and Ricci
tensor square one, this theory has drawn renewed interests in three
dimensional gravity in various view points, one of which is along
$AdS/CFT$ correspondence. In particular, it was shown that the
combination of curvature squared terms is consistent with and, in
fact, determined by the holographic c-theorem, which is a specific
incarnation of $AdS/CFT$
correspondence~\cite{Sinha:2010}$\sim$\cite{Liu:2009kc}. This
insight leads to the extensions of NMG to even higher curvature
theories. Other directions of the exploration of NMG include various
(charged) black hole
solutions~\cite{Bergshoeff:2009aq}\cite{Clement:2009gq}$\sim$\cite{Ghodsi:2010},
supersymmetric
extension~\cite{Andringa:2009yc}\cite{Bergshoeff:2010mf},
generalization by including the gravitational Chern-Simons
term~\cite{Liu:2009pha}, the computation of correlations and anomaly
via $AdS/CFT$ correspondence~\cite{Grumiller:2009sn}, the
appropriate extension of Gibbons-Hawking boundary
term~\cite{Hohm:2010jc}, the Hamiltonian analysis of
NMG~\cite{Blagojevic:2010ir}, etc.

In this paper, we consider the simplest version of NMG of which
action is given by\footnote{We have introduced  $\eta$ for  the
various sign choice of  terms in the action.}
\begin{equation}\label{NMG}
S =\frac{\eta}{2\kappa^2}\int d^3x\sqrt{-g}\bigg[ \sigma R +
\frac{2}{l^2} + \frac{1}{m^2}K  \bigg]\,,
\end{equation}
where $\eta$ and $\sigma$ take $1$ or $-1$, and $K$ is a specific
combination of scalar curvature square  and Ricci tensor square
defined by \beq
 K = R_{\mu\nu}R^{\mu\nu} -\frac{3}{8}R^2\,.
\eeq
Our convention is such that $m^2$ is always positive but the
cosmological constant $l^2$ has no such restriction. The equations
of motion(EOM)  of NMG are given by
\begin{equation}
\CE_{\mu\nu} =\eta\Big[
 \sigma G_{\mu\nu} - \frac{1}{l^2}g_{\mu\nu} + \frac{1}{2m^2}K_{\mu\nu}\Big]
        =0\,,
\end{equation}
 where
\begin{equation}
 K_{\mu\nu} = g_{\mu\nu}\Big(3R_{\alpha\beta}R^{\alpha\beta}-\frac{13}{8}R^2\Big)
                + \frac{9}{2}RR_{\mu\nu} -8R_{\mu\alpha}R^{\alpha}_{\nu}
                + \half\Big(4\nabla^2R_{\mu\nu}-\nabla_{\mu}\nabla_{\nu}R
                -g_{\mu\nu}\nabla^2R\Big)\,.
\end{equation}
There is an interesting relation between the scalar $K$ and the
tensor $K_{\mu\nu}$
\begin{equation}
 g^{\mu\nu}K_{\mu\nu} = K \,.
\end{equation}

Another useful form of the action which is equivalent to (\ref{NMG})
is given by
\begin{equation}
 S = \frac{\eta}{2\kappa^2}\int d^3x\sqrt{-g} \bigg[ \sigma R +
     \frac{2}{l^2} + f^{\mu\nu}G_{\mu\nu} - \frac{1}{4}m^2
     (f^{\mu\nu}f_{\mu\nu} - f^2) \bigg] \,,
\end{equation}
where $f_{\mu\nu}$ is an auxiliary symmetric tensor field related to
`Shouten' tensor~\cite{Gover:2008sw}
\begin{equation}
 f_{\mu\nu} = \frac{2}{m^2}S_{\mu\nu},~~~ S_{\mu\nu} \equiv
 R_{\mu\nu} - \frac{1}{4}Rg_{\mu\nu}
\end{equation}
and its trace $f = g^{\mu\nu}f_{\mu\nu}$.

\subsection{New type black holes}

Among   various black hole solutions known in NMG: BTZ black holes,
Warped $AdS_3$ black holes, New type of black holes, Lifshitz type
black holes,
etc~\cite{Bergshoeff:2009hq}\cite{Clement:2009gq}\cite{giribet:2009},
our main interests in this paper are the so-called (static) new type
of black holes which exist only when $\sigma=1$ and $l^2=1/m^2$.
These black hole solutions contain two parameters though the nature
of these parameters is obscure at the present. One of the
motivations for our study is to improve this situation. The metric
of new type black holes is given by
\beq\label{nmg-metric}
  ds^2 = L^2 \left[ -(r^2 + b r + c)dt^2 + \frac{dr^2}{(r^2 + b r + c)} + r^2 d\phi^2
  \right] \,,  \qquad L^2 = \frac{1}{2m^2}=\frac{l^2}{2}\,,
\eeq
with  outer and inner horizons  at $r_{\pm} = \half (-b \pm
\sqrt{b^2-4c})$.
The scalar curvature of this metric is given by
\[ R = - \frac{6}{L^2} - \frac{2b}{L^2r}\,,\]
which shows us that there is a curvature singularity at $r=0$ when
$b\neq 0$. New type black holes may be classified according to signs
of parameters $b$ and $c$.  This classification may be tabulated by
introducing a new parameter $q \equiv r_- / r_+$, as follows:  The
other sign combinations of $b$ and $c$ do not represent black holes.
\begin{center}
\begin{tabular}  [hb] { | c | c | c | }
\hline
   $q \equiv r_-/  r_+ $   &  $b$  and $c$    & $r_+$ and $r_-$   \\
 \hline
 $q=1$  & $ b<0,~  c>0$   ($b^2 =4c$ ) &  ${r_+}>0 ~,{r_-}>0$ $( {r_+}  = { r_-}  )$   \\
\hline
   $0<q<1$  & $b<0,~ c>0$    &  ${r_+}>0 ~,{r_-}>0$  $( {r_+} > { r_-} )$  \\
\hline
   $q=0$  &  $b<0,~ c= 0$   &$ r_+ > 0,~ r_-=0$   \\
 \hline
   $-1<q<0$  & $b<0,~ c<0 $   & $r_+ >0,~ r_-<0$  $(  {r_+}  >| { r_-} | )$ \\
 \hline
   $q = -1$  & $b=0,~ c<0$   & $r_+ >0, ~r_-<0$  $(  {r_+} =- { r_-} )$ \\
\hline
   $q< -1$ & $b>0,~ c<0$  &   $r_+ >0, ~r_-<0$  $(  {r_+}  < | { r_-} | )$ \\
\hline
\end{tabular}
\end{center}
 Note that new type black holes for  $q=-1$
 are nothing but  non-rotating BTZ black holes.

Now, let us list  some quantities of new type black holes. Firstly,
the Hawking temperature of new type black holes can be read from the
surface gravity or the  periodicity of Euclideanized action as
\begin{eqnarray}
 T_H = \frac{r_+ - r_-}{4\pi L} = \frac{\sqrt{b^2-4c}}{4\pi L} \,.
\end{eqnarray}
The mass of new type black holes is identified as a conserved charge by
\beq
M = \frac{b^2-4c}{16G} \,, \eeq
which is also justified by the dynamical approach or $AdS/CFT$
correspondence~\cite{Nam:2010dd}\cite{Nam:2010ma}\cite{Maeda:2010}.
Let us recall that the Bekenstein-Hawking-Wald entropy of new type
black holes is different from area law. More explicitly, it is given
by
\[ S_{BHW} = \frac{A_H}{4G}\eta\bigg[\sigma + \frac{1}{4m^2}\Big(R^{t}_{t} + R^{r}_{r}
-3R^{\phi}_{\phi}\Big)\bigg]_{\rm at~horizon}\,,
\]
where $A_H$ is the horizon area and $R^{\mu}_{\nu}$ are Ricci
tensors. By computing the entropy on the outer horizon,   this
entropy can be written as the difference between the area of the
outer horizon and the one of the inner horizon as follows
\beq S_{BHW}  = \frac{1}{4G}(A_+ - A_-) = \frac{\pi L}{2G}(r_+-r_-)
     = \frac{\pi L}{2G}\sqrt{b^2-4c} \,.
\eeq
It is interesting to note that the higher curvature effect  leads to
the inner horizon dependence of the black hole entropy.  The above
quantities satisfy the simple form of the first law of black hole
thermodynamics as $dM = T_HdS_{BHW}$, which is also consistent with
the $AdS/CFT$ correspondence.

Though some physical quantities are identified and thermodynamic
properties are checked,  the existence of two parameters is still
mysterious. It is unclear what is the meaning of  another
combination of parameters different from the mass combination, or what
combination at all one should take.   This problem may be rephrased
in another way as follows.

Usually higher derivative or curvature effects are regarded as
introducing additional degrees of freedom.  This is the reason that
the would-be nondynamical three dimensional gravity  allows
propagating massive graviton modes and becomes more interesting.  In
the form of initial value problems of EOM,  one needs more initial
values to solve the problem with higher derivatives.   In these
viewpoints, the above black hole entropy formula is
counterintuitive, since the positive inner horizon area means the
reduction of the entropy even if it comes from higher curvature
effects. More distinguishably, there is one parameter family of
black hole like solutions, that is, the extremal ones ($r_+ =r_-$),
which lead to zero entropy. Therefore, one may envisage the domain
of parameters of positive inner horizon area as forbidden or at
least as disconnected from the case of negative inner horizon area.
In a later section, we will show some clues on such expectation.

In order to understand these peculiar properties of the new type
black holes, it is desirable to investigate these black holes  in
more detail. In this paper we study scalar perturbations or scalar
fields on these black hole backgrounds. Specifically,  we focus on
the  QNMs associated with scalar fields. It turns out that QNMs give
some clues about the nature of parameters.

\section{Exact QNMs of New Type Black Holes}

The EOM of scalar fields on new type black hole backgrounds is
written as
\begin{equation}\label{wave-eq}
\nabla^2 \Psi - m^2 \Psi = 0\,,
\end{equation}
where $m$ is the mass of the scalar field. To obtain the QNMs of
scalar fields on the new type black hole background, we recall that
the metric of this black hole is given by (\ref{nmg-metric}). For
our convenience, let us take $L=1$ in the following, which can be
recovered by dimensional reasoning. By the separation of variables
$\Psi = R(r) e^{ i\omega t + i \mu \phi}$, one obtains the radial
equation of the the above EOM. Through the change of variable
\beq\label{chvar} z \equiv \frac{r-{r_+}}{r-{r_-}}\,, \eeq
it becomes
\begin{eqnarray}\label{eqradial}
&& R''(z) + \bigg[\frac{1}{z}-\frac{1}{z-1} +\frac{1}{z-z_0}
   \bigg] R'(z)  \nonumber \\
&& ~~~~~
   + {1 \over {z (z-1)(z-z_0)}}
   {\bigg[ \frac{ (z-1) (z-z_0)}{ z(z_0-1)^2 } {\w^2 \over {r_-^2}}
   -\frac{(z-z_0)}{(z-1)}  m^2  -\frac{(z-1)}{(z-z_0)}
   {\mu^2 \over {r_-^2}} \bigg] R(z)} =0  \,,
\end{eqnarray}
where $z_0 \equiv r_+/  r_- =1/q$ and $'$ denotes the
derivative with respect to  the variable $z$.

Defining $R(z)= z^\alpha (1-z)^\beta
(z_0-z)^\gamma \CH(z)$ and choosing parameters $\alpha$, $\beta$ and
$\gamma$  without loss of generality as
\begin{eqnarray}
 \alpha =  \frac{i \w }{(1-1/z_0){r_+}} \,,~~ \beta = 1 - \sqrt{1+m^2}
 ~~{\rm and}~~ \gamma=  \frac{\mu}{{r_+}}\sqrt{z_0} \,,
\end{eqnarray}
one obtains  a differential equation for $\CH(z)$ in the form of
\begin{eqnarray} \label{heun}
&& \CH''(z) + \bigg[\frac{2\alpha +1}{z}
   +\frac{2 \beta -1}{z-1}+\frac{2 \gamma+1}{z-z_0}\bigg] \CH'(z)
   + \bigg[z_0(\alpha+ \beta -2 \alpha \beta - \beta^2)   \nonumber \\
&& ~~~~~~~~~
   -\alpha -\gamma- 2 \alpha \gamma - \gamma^2 + \Big\{ (\alpha
   +\beta +\gamma)^2 -\alpha^2  \Big\} z  \bigg]
   {{\CH(z) } \over {{z (z-1)(z-z_0)}}} =0 \,,
\end{eqnarray}
which turns out to be  just a Heun's differential
equation~\cite{Heun}. Let us recall that  the Heun's  differential
equation in the standard form is given by
\[
  \CH''(z)+\bigg[\frac{\nu }{z}+\frac{\delta}{z-1}
   +\frac{\epsilon}{z-z_0}\bigg]  \CH'(z)
   + \frac{( \lambda  \xi z -\eta )}{z(z-1)(z-z_0)}  \CH(z) =0 \,,
\]
with the condition  $\epsilon=\lambda +\xi -\nu -\delta+1$. Now, one
can see that (\ref{heun}) is a Heun's differential equation by
identifying
\begin{eqnarray}
 && \nu= 2 \alpha+1 \,, ~~ \delta = 2 \beta-1 \,, ~~
    \epsilon= 2 \gamma+1 \,,
    \\
 && \eta = z_0(2\alpha\beta+\beta^2-\alpha-\beta)
    + \gamma^2+2\alpha\gamma+\alpha+\gamma\,,
    \label{lamcon1}\\
 && \{ \lambda \,,  \xi \}
    = \{ 2 \alpha + \beta + \gamma \,, ~ \beta + \gamma \}
    \label{lamcon}
    ~~ {\rm or} ~~ \{  \beta + \gamma \,,~
    2 \alpha + \beta + \gamma  \} \,.  \label{xicon}
\end{eqnarray}
There exists  a local solution of the differential equation
(\ref{heun}) near $z=0$ which is given by a combination of local
Heun function $H(z)$'s. By imposing ingoing boundary condition at
the outer horizon, one obtains  the radial function as
\beq R(z) =  z^\alpha (1-z)^\beta
(z_0-z)^\gamma
  {\cal H}(z) = C_1 z^\alpha (1-z)^\beta
(z_0-z)^\gamma    H(z_0,\eta, \lambda,\xi, \nu, \delta |z)\,. \eeq
Note that as $r$ goes to the horizon ($z\rightarrow 0$), $R(z)$ is
reduced to $z^\alpha$ which represents the ingoing mode since
$\alpha = \frac{i\w}{(1-1/z_0)r_+}$.

Though solutions for  the scalar field perturbation on new type
black holes can be written as a local Heun function, one cannot
extract useful information from it since Heun functions do not have
a closed form in most case and their properties are not well
understood, yet. Moreover, one should impose the boundary condition
at the spatial infinity for the complete specification of QNMs.
However, in the special cases of new type black holes, we can find
QNMs by reducing Heun functions to more familiar ones.

First of all, when $q=-1$, new type black holes are just the
non-rotating BTZ black holes and QNMs were already obtained  in
terms of hypergeometric functions, which can  also  be  derived in
our approach by the reduction of Heun functions to hypergeometric
ones.  When $q=0$, the radial function becomes a (reduced) confluent
Heun function. In this case  QNMs can also be obtained exactly using
the connection formula of the (reduced) confluent Heun functions.
When $q=1$, zero modes can be obtained. Now, let us elaborate on
these things.

Firstly, let us consider  the $q=-1$ case, i.e. $r_-  =-r_+$. The
radial function is given by
\beq \label{heunsol}
 R (z) =  C_1 z^\alpha (1-z)^\beta
(-1-z)^\gamma       H(-1,\eta,\lambda ,\xi ,2 \alpha +1,
       2 \beta-1 {\vert} z) \,, \eeq
where $\eta$ is given by Eq.~(\ref{lamcon1}) while $\{\lambda$,
$\xi\} $  is chosen as $\{ 2 \alpha + \beta + \gamma,  \beta +
\gamma \}$ without loss of generality. It has been  known that a
transformation from the Heun function to the hypergeometric function
exists when special relations among parameters are satisfied (see
Appendix A)~\cite{Heun-HyperG}. As one can check that parameters in
the case of $z_0=-1$ satisfy the conditions for the transformation,
the above solution in terms of Heun function (\ref{heunsol}), can be
written in terms of hypergeometric one. Using this transformation,
the solution of the radial equation can be written in the  form of
\beq
  R(z) = C'_1 z^{\alpha } (1-z)^{\beta }(1+z)^{-2\alpha-\beta}
     { _2F_1}\left(  \alpha +\frac{\beta +\gamma}{2}, \alpha
     +\frac{\beta -\gamma}{2},  2\alpha+1
     \Big| \frac{4z}{(z+1)^2}\right) \,.
\eeq
Inserting the values of $\alpha,\beta, \gamma$ with coordinate
change to $w=4z/(z+1)^2$, one can see that  this is the same
function already obtained for radial function on BTZ black
holes~\cite{Birmingham:2001}.  Therefore, imposing the boundary
condition (Dirichlet one) at the spatial infinity,  QNM frequencies
are obtained as follows:
\beq \label{qnmbtz} \w_{QNM}=\pm \mu + i 2 r_+ \Big[ n+{1 \over 2}
(1+\sqrt{1+m^2})\Big] \,,
\eeq
 where $n = 0, 1, 2, \cdots$.

Secondly, we consider the case $q=0$, i.e. $r_- = 0$.  This case can
be obtained by taking a limit $z_0 \rightarrow \infty$ with $r_- =
r_+/z_0$  in the Eq.(\ref{eqradial}). One can observe  that this is
the limit for the (reduced) confluent Heun's differential
equation~\cite{conf-Heun}. Therefore, Eq.(\ref{heun}) is  reduced to
a (reduced) confluent Heun's differential equation
\begin{eqnarray}
 {\cal H}''(z) +\left[\frac{ 2\bar{\alpha}+1}{z}
         + \frac{2\beta -1  }{z-1}\right] {\cal H}'(z)
         - \frac{(\bar{\gamma}^2 z-\bar{\eta})}{z(z-1)} {\cal H}(z)=0 \,,
\end{eqnarray}
 with the identification among
parameters
\begin{eqnarray}
 \bar{\alpha} \equiv \lim_{z_0\rightarrow \infty} \alpha = i\frac{\omega}{r_+}\,, \qquad
 \bar{\gamma}^2 \equiv  \lim_{z_0\rightarrow \infty} \frac{\gamma^2}{z_0}
 =\frac{\mu^2}{r_+^2}\,, \qquad
 \bar{\eta} \equiv \lim_{z_0\rightarrow \infty}  \frac{\eta}{z_0}
 = \bar{\gamma}^2  -\bar{\alpha} -\beta +2 \bar{\alpha} \beta +\beta^2 \,.
\end{eqnarray}
The local solution of this equation near $z=0$ is given by a linear
combination of the (reduced) confluent Heun function $H_C(z)$'s.
Imposing ingoing boundary condition at the outer horizon, the radial
function can be represented by
\begin{eqnarray}
  R(z) = C_2 z^{\bar{\alpha}}(1-z)^{\beta}H_C  \bigg( 0, 2 \bar{\alpha}, 2 \beta-2,
          - \bar{\gamma}^2, (\beta -1)^2 +\bar{\gamma}^2 \Big| z \bigg) \,.
\end{eqnarray}
In order to investigate the behavior at infinity, we can use the
connection formula (see Appendix A) and obtain relevant part of the
radial function for imposing the boundary condition at the infinity
as
\begin{eqnarray}
 R(z) = C_2 { {\Gamma(2\bar{\alpha} +1) \Gamma(2-2 \beta)}
        \over {\Gamma(3-2 \beta +\zeta) \Gamma(2 \bar{\alpha} -\zeta)} }
        z^{\bar{\alpha} } (1-z)^{1-\beta }  H_C \bigg(0,2 \beta-2,
        2\bar{\alpha}, \bar{\gamma}^2,\left.(\beta -1)^2\right|1-z\bigg) \,,
\end{eqnarray}
where $\zeta$ is determined by
\begin{eqnarray}
 \zeta^2 +(3-2 \bar{\alpha} -2 \beta)\zeta+ \bar{\gamma}^2
 +\beta ^2-3 \bar{\alpha} +2 \bar{\alpha}  \beta -3 \beta +2 =0  \,.
\end{eqnarray}
The Dirichlet boundary condition at the spatial infinity ($z=1$)
requires that the coefficient given by gamma functions vanich. This
gives the following QNM frequencies:
\begin{eqnarray}\label{mode}
\w_{QNM}= i r_+\left[\frac{{\mu^2 \over {2 r_+^2} }+
   (n+\sqrt{1+m^2})
   (n+\sqrt{1+m^2}+1)}{2
   (n+\sqrt{1+m^2})+1} \right] \,.
\end{eqnarray}
When $n$ is very large, the asymptotic form of QNM frequencies is given by
\begin{equation}
\w_{QNM} \sim  i {{r_+} \over {2}}n \,.
\end{equation}

Finally, let us consider the extremal case $q=1$, i.e. $r_-=r_+$
with denoting the  horizon radius as  $r_0=r_-=r_+$. In this case
the change of variable in  the Eq.(\ref{chvar}) is meaningless.
Instead, let us set the variable, $z \equiv \frac{r-r_0}{r}$ and
denote $R(z)= z^\alpha (1-z)^\beta {\cal G}(z)$. Then, the radial
equation in terms of ${\cal G}(z)$ can be written in the form of
\begin{eqnarray}\label{anoeq}
 {\cal G}''(z)+\bigg(\frac{2 (\alpha+1)}{z}
   +\frac{1-2 \beta}{1-z}\bigg) {\cal G}'(z)+ {1 \over  {z (1-z)}}
   {\bigg[ \bigg(\frac{1-z}{z^3}\bigg) \frac{\w^2}{r_0^2}
   +\frac{\mu^2}{r_0^2}-\alpha ^2-\beta ^2
   -2\alpha  \beta\bigg] {\cal G}(z)}=0 \,,
\end{eqnarray}
by choosing parameters as
\begin{eqnarray*}
\alpha = -\frac{r_0 \pm \sqrt{r_0^2+4 \mu^2+4 m^2 r_0^2}}{2 r_0} \,,
~~ \beta =1 \pm \sqrt{1+m^2} \,.
\end{eqnarray*}
When $\w=0$, it becomes the hypergeometric differential equation. In
this case, one obtains the solution as
\begin{eqnarray}
 R(z) &=& c_1 z^\alpha (1-z)^\beta \, _2F_1\Big(-\frac{\mu}{r_0}
      +\alpha +\beta , \frac{\mu}{r_0}+\alpha +\beta,
      2 \alpha +2 {\Big|} z\Big) \nonumber\\
      && + c_2 \, z^{-\alpha -1}(1-z)^{\beta}\, _2F_1\Big(-\frac{\mu}{r_0}
      -\alpha +\beta -1,\frac{\mu}{r_0}-\alpha +\beta -1,
      -2 \alpha  {\Big|} z\Big) \,.
\end{eqnarray}

Considering another choice of change of variable and an appropriate
ansatz for radial function, e.g. $z'=\frac{r}{r-r_0}$ and
$R(z')=z'^{\alpha'}(1-z')^{\beta'}e^{\chi z'}G(z')$ , the radial
equation can be reduced to a confluent Heun's differential
equation\footnote{It is equivalent to the generalized spheroidal
wave equation~\cite{spheroidal-Eq}.}. Because of the absence for the
connection formula among the confluent Heun functions at
singularities, it is hard to find exact QNMs for this case. On the
other hand, there were some other attempts to find QNMs analytically
by perturbation method for Heun's differential
equation~\cite{Siopsis:2003}\cite{Siopsis:2003-1}.

\section{Asymptotic QNMs in New Type Black Holes}

In this section, we present the asymptotic QNM frequencies of
scalar fields on new type black holes. To obtain these, we  use the
so-called Stokes line method  which matches the approximate solutions of the radial
equation by  the analytic continuation of the radial
coordinate to the complex plane~\cite{motl:2003}. It is sufficient to consider
  the asymptotic solutions near the infinity, the origin and the
event horizon. After the approximate solutions in each relevant
region are obtained, they are matched with appropriate boundary
conditions. Through these matchings, one can read off  the QNM
frequencies of scalar fields on new type black holes.

Firstly, we need to solve the scalar perturbation equation
(\ref{wave-eq}) in each region. For this purpose it is convenient to
introduce the tortoise coordinates $x$ as follows
\begin{eqnarray}
  \frac{dx}{dr} = \frac{1}{(r-r_+)(r-r_-)}\,.
\end{eqnarray}
By choosing the integration constant such as $x(r=0)=0$, one obtains
\beq x  =  \frac{1}{r_+-r_-}\ln\frac{r-r_+}{r - r_-} +x_0\,,   \eeq
where   $x_0$ is given by
\beq\label{x-naught}
 x_0  \equiv \frac{1}{r_+ - r_-}\ln \Big( \frac{r_-}{r_+} \Big)
     = \frac{1}{2\kappa_+}\ln \Big( \frac{r_-}{r_+} \Big)\,.
\eeq
It is interesting to note that $r_+$ should  always be taken as a
positive number to form black holes whereas $r_-$ is allowed to be
any value less than  or equal to $r_+$. When $r_- \le 0$, there is
just one real horizon. In the viewpoint of asymptotic QNM
computation, $r_-=0$ is a special case and should be treated
separately, whereas the case of $r_- < 0$ can be done uniformly.
In each region, the tortoise coordinate $x$ behaves like
\begin{eqnarray*}
 x \simeq \left\{
              \begin{array}{ll}
                 \frac{r}{r_+ r_-}  & \qquad    r \rightarrow 0\\
                 -\infty   & \qquad    r \rightarrow r_+   \\
                 x_0 - \frac{1}{r}  & \qquad    r \rightarrow \infty
              \end{array}
          \right.  \,.
\end{eqnarray*}
By denoting $\Phi\big(x(r)\big) \equiv \sqrt{r} R(r)$, the radial
equation  of scalar field equation can be  written as
\begin{equation}\label{eq-toto}
  \bigg[-\frac{d^2}{dx^2} - \omega^2 +U(x)\bigg] \Phi(x) =0\,,
\end{equation}
where the potential term is given by
\begin{equation}
 U(x) =  \Big(1-\frac{r_+}{r}\Big)\Big(1-\frac{r_-}{r}\Big)
         \bigg[\mu^2 + m^2 r^2+ \frac{1}{4}
         \Big(3r^2 - (r_+ +r_-)r - r_+r_-\Big)\bigg] \,.
\end{equation}
The potential term in each region becomes
\begin{eqnarray}\label{potent}
 U(x) \simeq
      \left\{
          \begin{array}{ll}
            \frac{r_+ r_-}{r^2} \left( \mu^2 - \frac{r_+ r_-}{4} \right)
            = \frac{j^2-1}{4x^2}\,,  \qquad  j^2 \equiv \frac{4\mu^2}{r_+ r_-}
            &  \qquad  r \rightarrow 0 ~~(x \rightarrow 0)  \\ \nn \\
            0  & \qquad  r \rightarrow r_+  ~~(x \rightarrow -\infty)   \\ \nn \\
            \big(\frac{3}{4} + m^2 \big) r^2
            = \frac{j^2_{\infty}-1}{4(x-x_0)^2}  \,, \qquad
            j_{\infty} \equiv  2\sqrt{1+m^2}    & \qquad  r \rightarrow  \infty
            ~~(x \rightarrow x_0)
          \end{array}
      \right. .
\end{eqnarray}
In each  singular point region, one can obtain the radial function,
$\Phi(x)$, as follows
\begin{eqnarray}\label{sols}
 \Phi(x) = \left\{
      \begin{array}{ll}
      A_+ \sqrt{2\pi\omega x} J_{\frac{j}{2}}(\omega x)
      + A_- \sqrt{2\pi\omega x} J_{-\frac{j}{2}}(\omega x) \,,
      &  \qquad  r \rightarrow 0 ~~(x \rightarrow 0) \\   & \\
      C_+ \sqrt{2\pi\omega (x_0-x)}
      J_{\frac{j_{\infty}}{2}}(\omega(x_0-x))   &    \\
         ~~~~~~~~~~   + C_- \sqrt{2\pi\omega (x_0-x)}
      J_{\frac{-j_{\infty}}{2}}(\omega(x_0-x)) \,,
      & \qquad  r \rightarrow  \infty ~~(x \rightarrow x_0) \\  & \\
      D_+ e^{i\omega x}  + D_- e^{-i\omega x}\,,
      & \qquad  r \rightarrow r_+  ~~(x \rightarrow -\infty)
      \end{array}
     \right.
\end{eqnarray}
where $A_\pm$,$C_\pm$ and $D_\pm$ are complex constants and
$J_{\pm\frac{j}{2}}(\w x)$ and
$J_{\pm\frac{j_{\infty}}{2}}(\w(x_0-x))$ represent first kind Bessel
functions. The Dirichlet boundary condition near the spatial
infinity implies that this radial function should vanish there.
Therefore we should take $C_- = 0$.   The condition of  $D_- =0$
comes from the boundary condition for QNMs near the event horizon
$r_+$: $\Phi(x)$ should contain only ingoing mode $e^{i\omega x}$
there.

In order to obtain asymptotic QNM frequencies, let us  examine the
Stokes line which is defined by the curve ${\frak I}m (\omega x) =
0$. For the asymptotic $AdS_D$ solutions ($D \ge4$), it has been
known that  the magnitudes of both ${\frak I}m(\omega x)$ and
${\frak R}e(\omega x)$ are comparable in the asymptotic QNM
frequencies, which is related to the fact  that $\omega x$ is
asymptotically real when $x$ goes to $x_0$, i.e. $r \rightarrow
\infty$. Along this Stokes line, modes will be oscillatory without
any exponentially growing/decaying modes. In a neighborhood of the
origin, the relation $x \sim \frac{r}{r_+ r_-}$ shows us that
\begin{equation*}
 r = \rho e^{i(n\pi + \theta_0)} \,,
\end{equation*}
where $\rho > 0$ and $n = 0, 1$. It describes two lines emanating
from the origin, spaced by an angle $\pi$. Note that the sign of
$\omega x$ on these lines is alternatively positive and negative.
The angle $\theta_0$ can be obtained by
\begin{equation}
  \theta_0 = {\rm arg}(x_0)
    = \left\{\ba{ll} 0  & \qquad {\rm for}~ r_- >0\,, \\
    \tan^{-1}\Big(\frac{\pi}{\ln |\frac{r_-}{r_+}|}\Big)
    & \qquad {\rm for}~ r_- <0\,.\ea\right.
\end{equation}
In the case of  $r_- < 0$,   the branch of the Stokes line emanating
from the origin with the angle $\theta_0$ goes to the infinity with
positive $\omega x$ value. The other branch runs to the horizon with
negative $\omega x$ value starting from the origin with its angle
$\theta_0-\pi$. The educated guess leads to  Stokes line  depicted
in the Figure 1, which results in QNM frequencies consistent with
numerical ones given in the Appendix B. In $r_- >0$ case, the Stokes
lines seems to be laid on the real axis. This lines do not seem to
make sense in matching  solutions around singular points. Therefore,
we confine ourselves to the parameter domain $r_- <0$ case only.
%

\begin{figure}  [hbtp]
\centering
\includegraphics[width=9cm,height=7cm]{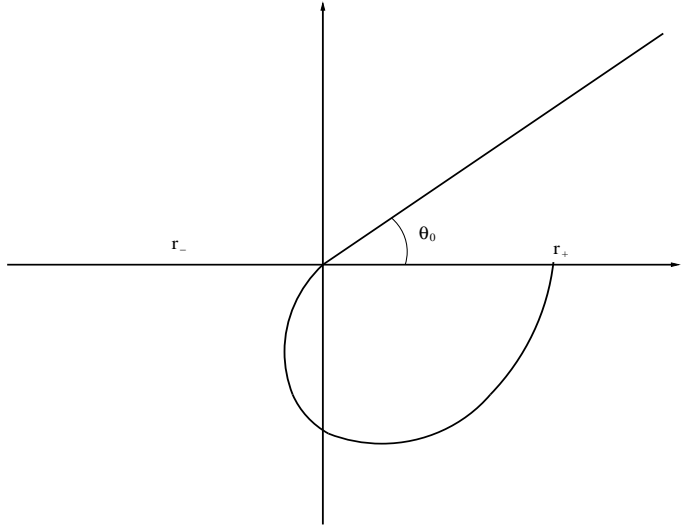}
\caption{Stokes line}
\end{figure}

Now, we need to match the above solutions along the stokes line on
the complex $r$ plane. Using the asymptotic expansion of the first
kind Bessel function in the region of $|\omega x| \rightarrow
\infty$~\cite{Ambra},  one obtains the expansions of the first and
second solutions of Eq.(\ref{sols}) at some point on the positive
branch, respectively as
\begin{eqnarray}
 \Phi_{o}(x) &\sim & e^{-i\frac{\pi}{4}}\big( A_+ e^{-i\frac{j\pi}{4}}
         + A_- e^{i\frac{j\pi}{4}} \big)e^{i\omega x}
         + e^{i\frac{\pi}{4}}\big( A_+ e^{i\frac{j\pi}{4}}
         + A_- e^{-i\frac{j\pi}{4}} \big)e^{-i\omega x}\,,
         \label{match-1} \\ \nn \\
 \Phi_{\infty}(x) &\sim & e^{i\frac{\pi}{4}}\big( C_+ e^{-i\omega x_0}
         e^{i\frac{j_{\infty}\pi}{4}} \big) e^{i\omega x}
         + e^{-i\frac{\pi}{4}}\big( C_+ e^{i\omega x_0}
         e^{-i\frac{j_{\infty}\pi}{4}} \big) e^{-i\omega x} \,.
         \label{match-2}
\end{eqnarray}
Matching these two functions gives us a relation
\begin{eqnarray}\label{rel-1}
 A_+ \cos \bigg[ \omega x_0 - \frac{(j_{\infty}+j)\pi}{4} \bigg]
     + A_- \cos \bigg[ \omega x_0 - \frac{(j_{\infty}-j)\pi}{4}
     \bigg]
     = 0  \,.
\end{eqnarray}

The solution near the origin (\ref{match-1}) is on the positive
branch of the Stokes line but the solution near the horizon is on
the negative one. In order to match the near-horizon solution with
the near-horizon one, one performs the rotation $x'=e^{-i\pi}x$ in
the neighborhood of the origin while evading the singular origin. By
doing this rotation near the origin and then using asymptotic
expansion for the first solution of Eq.(\ref{sols}), the solution
$\Phi_o(x)$ on the negative branch is represented by
\begin{eqnarray}
 \Phi_o(x) \sim e^{-i\frac{\pi}{4}} \big( A_+ e^{-i\frac{j\pi}{4}}
     + A_- e^{i\frac{j\pi}{4}} \big) e^{i\omega x}
     + e^{-i\frac{3\pi}{4}}\big( A_+ e^{-i\frac{3j\pi}{4}}
     + A_- e^{i\frac{3j\pi}{4}} \big) e^{-i\omega x} \,.
\end{eqnarray}
By matching the above solution with the solution near  the event
horizon given in the Eq.(\ref{sols}) one can see  that $\Phi(x)$
should be given  in the form of $e^{i\omega x}$. Consequently one
obtains another relation
\begin{eqnarray}\label{rel-2}
 A_+ e^{-i\frac{3j\pi}{4}} + A_- e^{i\frac{3j\pi}{4}} = 0 \,.
\end{eqnarray}
The existence of nontrivial solutions of two relations (\ref{rel-1})
and (\ref{rel-2}) on the coefficients  $A_+$ and $A_-$ implies that
their determinant should vanish.  As a result,  one obtains
\begin{eqnarray}
 \omega x_0 = \Big( n+\frac{1}{2}+\frac{j_{\infty}}{4} \Big)\pi
      -\frac{i}{2}\ln \Big[ 2\cos\big(\frac{j\pi}{2}\big) \Big] \,.
\end{eqnarray}
Note that $q$ is negative since we are dealing with the case
$r_-<0$.

Since we know the exact form of $x_0$ given in the Eq.
(\ref{x-naught}), we can obtain the QNM frequencies analytically.
Substituting $\kappa_+ = (1-q)r_+/2 = 2\pi T_H$ into
Eq.~(\ref{x-naught}), we obtain
\begin{eqnarray}
 x_0 = \frac{1}{4\pi T_H} \ln (q) = \frac{1}{4T_H} \Big(i +
       \frac{1}{\pi} \ln |q| \Big)\,.
\end{eqnarray}
%
%
%
Rewriting $j$ in terms of $q$ as
\begin{eqnarray}
 j  = i\frac{1-q}{\sqrt{|q|}}\frac{\mu}{2\pi T_H} \,.
\end{eqnarray}
QNM frequencies of scalar
perturbation on new type black holes can be written as
\begin{eqnarray}
 \omega_{QNM}  &=&
       i \frac{4\pi^3 T_H}{\pi^2+(\ln|q|)^2}
       \Big( n+\frac{1}{2}+\frac{1}{2}
       \sqrt{1+m^2} \Big)
       + i\frac{2\pi T_H \ln|q|}{\pi^2 + (\ln|q|)^2}
       \ln \Big[ 2\cosh \Big(\frac{1-q}{\sqrt{|q|}}
       \frac{\mu}{4T_H}\Big) \Big]        \nonumber \\   \\
    && - \frac{4\pi^2 T_H \ln|q|}{\pi^2+(\ln|q|)^2}
       \Big( n+\frac{1}{2}+\frac{1}{2}
       \sqrt{1+m^2} \Big)
       + \frac{2\pi^2 T_H}{\pi^2+(\ln|q|)^2}
       \ln \Big[ 2\cosh \Big(\frac{1-q}{\sqrt{|q|}}
       \frac{\mu}{4T_H}\Big) \Big] \,. \nn
\end{eqnarray}
In general, the above QNM frequencies show us that the dependence of
overtone number $n$ appears in both real and imaginary parts. So
$\frak{I}m (\omega)$ and $\frak{R}e (\omega)$ take comparably large
values with large $n$. This is consistent with the numerical results
in large $|r_-|$ value region,  some of which are represented in the
appendix B. From the above formula, one can see that the QNM
frequencies are  linear functions of temperature $T_H \sim \sqrt{M}$
for small $\mu$ values, which tells us some information of black
holes: At least for small $|q|$, the overtone number dependence
indicates that $M\sim (r_+-r_-)^2$.

In the limit of the non rotating BTZ black hole case i.e. $q=-1$ QNM
frequencies become the simple form of
\begin{eqnarray}
 \omega_{QNM} = i 4\pi T_H \Big( n+\frac{1}{2}+\frac{1}{2}
       \sqrt{1+m^2} \Big)
       + 2T_H \ln \Big[ 2\cosh \Big(\frac{\mu}{2T_H}\Big) \Big] \,.
\end{eqnarray}
The exact result of QNM frequencies has appeared
in~\cite{Birmingham:2001}, which is given by the Eq.(\ref{qnmbtz})
in the previous section. For the small value of $\mu$ and the large
value of the overtone number $n$ ($|\mu| \ll n$), the real part of
the QNM frequencies from the Stokes line method approaches to $2T_H
\ln2$ which is much smaller than the imaginary part, $4\pi n T_H$.
Therefore we may regard the QNM frequencies of the BTZ black hole as
a nearly pure imaginary number which is consistent with the
asymptotically reality condition $\omega x_0$ for small $\mu$ and
large $n$.

\section{Summary and Conclusion}

In this paper  we obtained QNM frequencies of scalar perturbation on
static new type black holes in NMG. Though  scalar field equations
were solved exactly in terms of Heun function which is the solution
of differential equations with  four regular singular points,  one
cannot obtain much information from this representation since most
properties of Heun functions are yet unknown. Specifically,  the
so-called connection formulae among  singular points are essential
for the purpose of obtaining QNM frequencies by imposing boundary
conditions at different singular points. Nevertheless, in some
special cases, for instance in the BTZ limit, one can reduce Heun
functions to more tractable ones and then obtain exact QNM
frequencies analytically. However, since such reduction does not
happen in most cases, we have adopted the Stokes line method to
obtain asymptotic QNM frequencies analytically, which is supported
by numerical results for the parameter domain $ -r_- > r_+$.

These results reconfirm the mass formula of new type black holes and
give some clues about their additional parameter. A  natural
combination of black hole parameters appears in $\omega_{QNM}$ is
the form of $Q\equiv \ln|q| = \ln |r_-/r_+|$. Moreover,  our results
for small $Q\equiv \ln|q|$  leads to $\omega_{QNM} \sim T_H n \sim n
(r_+ - r_-)$,  which   implies  $M\sim (r_+-r_-)^2$, since  the
overtone number, $n$, dependence of  the imaginary part of
asymptotic QNM frequencies is usually given by the mass of black
hole backgrounds, or  Choptuik scaling parameter.  This  is
consistent with results from other approaches.    Though this does
not  reveal the nature of additional parameter completely, it gives
some clues about its nature.  As alluded in the introduction, we
have showed that there seems disconnection between $r_- < 0$ and
$r_- >0$ during our analysis. That is to say, Stokes line method or
numerical one breaks down when $r_-=0$. More explicitly,  two
methods are inapplicable in this case because $x_0$ cannot be
defined  and the potential of the  radial equation in Eddington
coordinates  is unbounded.  This may not be just inadequacy of
methods to the problem as was argued by the entropy reduction in the
$r_- >0$ domain.

There are still many issues to be addressed in the future. First of
all,  it is very desirable to study $AdS/CFT$ correspondence for QNM
frequencies in new type black holes.  Since new type black holes
have different metric fall-off tails at the spatial infinity from
BTZ black holes, the dual CFT may be different. Nevertheless, it is
still perplexing that there is an additional parameter  in the dual
CFT picture. Thermal two point functions of perturbation operators
in CFT can be determined more or less uniquely only by conformal
symmetry and depends only on the conformal weights, which have been
identified as black hole mass or temperature. It is unclear how to
identify the additional parameter in the dual CFT.  Maybe new type
black holes are not good quantum objects allowed in quantum gravity.

Secondly, though we presented  exact QNM frequencies for $r_-=0$ and
$r_- = -r_+$, these cases should be studied more carefully, since
the potential in tortoise coordinates can have negative values when
$0 <  -r_- < r_+$ or $r_- \ge 0$. This can be referred to as the
necessity of further study about stability of new type black holes.
During our analysis, we have assumed that new type black holes are
stable under perturbations. However, this is unverified assumption,
though it is very plausible that new type black holes are stable
near the parameter region corresponding to BTZ black holes which are
stable since those can be embedded as Bogomol'nyi-Prasad-Sommerfield
(BPS) states in the supersymmetric extension of NMG.

Thirdly, it is worthwhile to consider the problem on the
quantization of the black hole entropy with the asymptotic QNMs.
There have been many investigations concerned with this
~\cite{Hod:1998}\cite{magi:2007}\cite{Kunstatter}$\sim$\cite{Kwon:2010-1}.
According to the work of \cite{Kwon:2010}\cite{Kwon:2010-1}, we can
calculate the entropy spectra of new type black holes from
asymptotic QNMs (i.e, for large overtone number). When $q= -1$ and
$q=0$, the entropy spectrum seems to be equally spaced. However,
when $ q<0 (q \neq -1)$, the spacing of the entropy spectrum depends
on the $q$ factor, even though the entropy spectrum is still equally
spaced for fixed $q$ value. The different aspect of the entropy
spectrum may be caused by the different behavior of the asymptotic
QNMs with both highly oscillating and damping modes. Some similar
phenomena are observed for large Schwarzschild-AdS black holes in
$D\ge 4$. Further investigations are needed to understand the
entropy spectrum using the asymptotic QNMs.

Finally, though numerical results support Stokes line or monodromy
method in general, the rigorous mathematical justification of this
method is still lacking. More concretely, it is unclear how to
estimate errors in Stokes line method and how to systematically
compute the next orders in this approach. Usual convergent or
asymptotic series approach to QNM frequencies has definite error
estimates and can be investigated systematically, which leads to
numerical computation of QNM frequencies.  Therefore, it is also
very interesting to investigate Stokes line method rigorously and
verify its validity in  general.

\section*{\center Acknowledgements}

S.H.Y was supported by the National Research Foundation of
Korea(NRF) grant funded by the Korea government(MEST) through the
Center for Quantum Spacetime(CQUeST) of Sogang University with grant
number 2005-0049409 and in part with the grant number 2009-0085995.
S.H.Y would like to thank Prof. Seungjoon Hyun at Yonsei University
for discussion some time ago. This work of S.N and J.D.P was
supported by a grant from the Kyung Hee University in
2009(KHU-20110060). S.N and Y.K were supported by Basic Science
Research Program through the National Research Foundation of
Korea(NRF) funded by the Ministry of Education, Science and
Technology(No.2010-0008109). S.N was also supported by the National
Research Foundation of Korea(NRF) grant funded by the Korean
government(MEST)(No. 2009-0063068).

\newpage
\appendix

\section{The reduction of the Heun function to the Hypergeometric function}
In this appendix  some mathematical identities are presented, which
are useful in the main text.
\subsection{Heun to hypergeometric}
The reduction formulae from the Heun function to the Hypergeometric
one  by polynomial transformations are given in mathematical
literatures, for instance see~\cite{Heun-HyperG}.  In summary,  when
four singular points, $(0,1,z_0,\infty)$ of the Heun function
satisfy harmonic or equianharmonic conditions there are polynomial
transformation from Heun function to Hypergeometric one. The
harmonic condition means the three points $(0,1,z_0$) are collinear
and equally spaced, which is relevant to our case.
In~\cite{Heun-HyperG}, $z_0=2$ is taken as the canonical value with
the polynomial transformation $R(t) = t(2-t)$, see the Eq.~(3.5.a).
The reduction formula in the case of  $z_0=-1$ can be obtained by
choosing   a M\"{o}bius transformation from $z$ to $t$  as
\[  t  \equiv \frac{2z}{z+1}\,. \]
Noting   that $R(t) = t(2-t) = \frac{4z}{(z+1)^2}$  and  (3.5.a)
actually means two equations, one can see that  there are two
possible ways to reduce Heun functions to Hypergeometric ones as
follows. That is
\beq \label{appent}
 H(-1,\eta,\lambda, \xi,\nu, \delta \, |\, z)
 = (z+1)^{-\alpha}\, {_2 F_1} \bigg(\frac{\lambda}{2},\frac{\nu
 +\delta-\xi}{2},\gamma \,\Big|\, \frac{4z}{(z+1)^2}\bigg)\,,
\eeq
with  relations among parameters  as
\beq \label{appent1} \eta = \lambda (\xi - \delta)\,, \qquad
\gamma=1+\lambda-\xi \,,\eeq
or the similar formula with $\lambda$ and   $\xi$ interchanged. This
formula explains the reduction of radial equation from new type
black holes to BTZ ones when $r_- = - r_+.$ Here, our convention of
the Hypergeometric function is
\beq 0 = \bigg[\frac{d^2}{dw^2} + \Big(\frac{c}{w} + \frac{1+a+b-c
}{w-1} \Big)\frac{d}{d w}- \frac{a b}{w(1-w)}\bigg]\,
{_2F_1}(a,b,c\, |\, w    )\,.    \eeq
%

 \subsection{Confluent Heun to hypergeometric}
%
The (reduced) confluent Heun's differential equation is given by
\begin{eqnarray}
  {\cal H}''(z) +\left(\frac{b+1}{z}+\frac{c+1 }{z-1}\right) {\cal H}'(z)
       + \frac{( d z -\epsilon)}{z(z-1)} {\cal H}(z)=0 \,.
\end{eqnarray}
The general solution of this equation near $z=0$ is given by a
linear combination of the (reduced) confluent Heun function $H_C
(z)$'s
\begin{eqnarray}
 {\cal H} (z) = c_1  H_C ( 0, b, c, d, e  | z )
      +  c_2 z^{-b} H_C ( 0, -b, c, d, e  | z )
\end{eqnarray}
where $e \equiv   \frac{1}{2} \{1-(b+1) (c +1)-2 \epsilon\} $.

There is a connection formula which is given by
\begin{eqnarray} \label{heunctran}
 H_C ( 0, b, c, d, e  | z )
      &=& c_1 { {\Gamma(b+1) \Gamma(-c)}
          \over {\Gamma(1-c+\zeta)\Gamma(b-\zeta)}}
          H_C ( 0, c, b, -d, e+d | 1-z )
          \nonumber \\
      &&  + c_2 { {\Gamma(b+1) \Gamma(c)}
          \over {\Gamma(1+c+\sigma)\Gamma(b-\sigma)}}
          (1-z)^{-c} H_C ( 0, - c, b, -d, e+d | 1-z ) \,,
\end{eqnarray}
where
\begin{eqnarray*}
 && \zeta^2 +(1-b-c)\zeta
     -\epsilon -b-c+{ d \over 2} =0 \,,\\
 && \sigma^2 +(1-b+c)\sigma
     - \epsilon -b(c+1)+{ d \over 2} =0 \,.
\end{eqnarray*}
The above formula means that the confluent Heun function about $z=0$
can be connected with some combination of the two solutions about
$z=1$ by analytic continuation.

When $d=0$, the reduction formula from the (reduced) confluent Heun
function to the hypergeometric one is given  as follows:
\begin{eqnarray} \label{ctohy}
 H_C\left( 0, b, c, 0, e  | z \right)
    &=& {_2 F_1} \left(A, B, C | z \right ) \,,
\end{eqnarray}
where
\begin{eqnarray*}
   \{ A, B \} &\equiv&  \left\{
       {{b+c \pm \sqrt{b^2+c^2-4 e+1}+1} \over 2},
       {{b+c \mp \sqrt{b^2+c^2-4 e+1}+1} \over 2}  \right\}  \nonumber \\
       C &\equiv & b+1  \,.
\end{eqnarray*}

 \subsection{Hypergeometric function and connection formula }
The standard form of the hypergeometric differential equation is
given by
\begin{eqnarray}
 F''(z)+{ {c-(a+b+1) z} \over {z (1-z)} } F'(z)
   - { {a b} \over  {z (1-z)} } F(z) =0  \,.
\end{eqnarray}
The solutions near $z=0$ is given by
\begin{eqnarray}
F(z) =  {_2F_1}(a,b,c |z)  + z^{1-c} {_2F_1}(a-c+1,b-c+1,2-c |z) \,.
\end{eqnarray}
The connection formula from the hypergeometric function,
${_2F_1}(a,b,c |z)$, about $z=0$ to another one about $z=1$ is given
by
\begin{eqnarray}
 {_2F_1}(a,b,c |z)
&=&  (1-z)^{c-a-b} \frac{\Gamma (a+b-c) \Gamma (c)
   }{\Gamma (a) \Gamma
   (b)} {_2F_1}{(c-a,c-b, 1 -a-b+c | 1-z)}  \nonumber \\
&& +   \frac{ \Gamma (c) \Gamma
   (-a-b+c)}{\Gamma (c-a) \Gamma (c-b)} {_2F_1}{(a,b,1+a+b-c | 1-z)}
   \,.
\end{eqnarray}
%

\section{Numerical results}

In this appendix, we present some details about the numerical
computation of QNM frequencies
following~\cite{Horo:99}\cite{Cardoso:2001}\cite{Cardoso:2004}. New
type black holes with ingoing Eddington coordinates($v\equiv t+x$)
are given by
\beq ds^2 = L^2\bigg[ -f(r)dv^2 + 2dv dr + r^2d\phi^2\bigg]\,,
\qquad f(r) \equiv (r-r_+)(r-r_-)\,. \eeq
Scalar fields on these black hole backgrounds may be decomposed as
$\psi = \frac{1}{\sqrt{r}} \Phi(r)\,  e^{i\omega v} e^{i \mu\phi}$
under the convention taken in this paper. Note that in this
convention $\omega_{QNM}$ should have positive imaginary part. The
above separation of variables leads to the radial equation of
massless scalars as
\beq  f(r)\Phi''(r) + \Big(f'(r) +2i\omega\Big)\Phi'(r) -V(r)\Phi(r)
=0\,, \qquad '\equiv \frac{d}{dr}\,, \eeq
where
\bea V(r) &\equiv& -\frac{1}{4r^2}f(r) + \frac{1}{2r}f'(r) + \frac{\mu^2}{r^2}  \nn \\
&=& \frac{3}{4}-\frac{r_++r_-}{4r} + \frac{\mu^2-r_+r_-}{4r^2}\,.
\nn \eea

By introducing new variable $y \equiv 1/r$, one obtains
\[ s(y)\frac{d^2}{dy^2}\Phi + \frac{t(y)}{y-y_+}\frac{d}{dy}\Phi + \frac{u(y)}{(y-y_+)^2}\Phi =0\,, \]
where $y_+ \equiv 1/r_+$, $y_-\equiv 1/r_-$ and
\bea s(y) &\equiv&  -\frac{y^2}{y_+y_-}(y-y_-)\,, \nn \\
        t(y) &\equiv&  -\frac{2}{y_+y_-}y(y-y_+)(y-y_-)
             +  2y - \frac{y_+ + y_-}{y_+y_-}y^2 +2i\omega  y^2\,, \nn \\
        u(y) &\equiv& (y-y_+)V(r)\,. \nn \eea
Series solution ansatz is taken as
\[ \Phi(y) = \sum_{\ell=0}^{\infty}a_\ell(y-y_+)^2\,, \]
which gives us recursion relation of $a_\ell$ in terms of the
coefficients of mode expansion of $s(y),t(y),u(y)$
\bea a_\ell &=& - \frac{1}{P_\ell}\sum_{k=0}^{\ell-1}
      \Big[k(k-1)s_{\ell-k}+kt_{\ell-k}+u_{\ell-k}\Big]a_k\,, \nn \\
  P_\ell &=& \ell (\ell-1)s_0+\ell t_0
      = 2y^2_+\ell(\ell\kappa_+ + i\omega)\,. \nn \eea
These formulae can be implemented easily in
mathematica~\cite{mathe}, which leads to the following results in the case of $ - r_-  \ge  r_+$ as follows.
\begin{center}
\begin{tabular}{ | c | c | c | c | } \hline
$r_+$   & $r_-$  & $\mu$ & $~~~\omega_{QNM}$ \\ \hline 1 & -10 & 0 &
18.0486 + 20.9188 i \\ \hline 1 & -10 & 1 &  18.5137+ 20.5174 i \\
\hline 1 & -10 & 2 & 19.5422+ 19.6755 i  \\      \hline 1 & -10 & 3
& 20.7523+ 18.7317 i \\    \hline 1 & -10 & 4 & 22.0311+ 17.7672 i
\\    \hline 1 & -10 & 5 &  23.3505+ 16.7959 i\\  \hline
\end{tabular}
~~
\begin{tabular}{ | c | c | c | c | } \hline
$r_+$   & $r_-$  & $\mu$ & ~~~$ \omega_{QNM}$ \\   \hline 100 & -200
& 0 &   108.272 + 284.046  i \\     \hline 100 & -200 & 1 & 108.282+
284.044 i\\     \hline 100 & -200 & 2 &    108.312+ 284.036     i
\\   \hline 100 & -200 & 3 &  108.363 + 284.024   i \\ \hline  100 &
-200 & 4 &   108.433 + 284.006   i \\    \hline 100 & -200 & 5 &
108.524 + 283.984   i
\\    \hline
\end{tabular}
\end{center}

\begin{center}
\begin{tabular}{ | c | c | c | c | } \hline
$r_+$   & $r_-$  & $\mu$ & ~~~$ \omega_{QNM}$ \\   \hline 10 & -20&
0 &   10.8272 + 28.4046     i \\     \hline 10 & -20 & 1 & 10.9274 +
28.3797    i \\     \hline 10 & -20 & 2 &  11.2203 + 28.3069    i \\
\hline 10 & -10 & 3 &    11.6854 + 28.1916      i \\    \hline 10 &
-10 & 4 &    12.2947 + 28.0413     i \\    \hline 10 & -10 & 5 &
13.0192 + 27.8644   i
\\    \hline
\end{tabular}
~~
\begin{tabular}{ | c | c | c | c | } \hline
$r_+$   & $r_-$  & $\mu$ & ~~~$ \omega_{QNM}$ \\   \hline 10 & -10&
0 &   0.00000+ 20.0044  i\\     \hline
10 & -10 & 1 &   1.00022 + 20.0022    i \\
\hline 10 & -10 & 2 & 2.00045+ 20.0022   i\\   \hline 10 & -10 & 3 &
3.00068 + 20.0022    i \\    \hline 10 & -10 & 4 &   4.00091 +
20.0022    i \\    \hline 10 & -10 & 5 &  5.00113 + 20.0022  i
\\    \hline
\end{tabular}
\end{center}
One can see that there are very small variations of the numerical
results when $\mu$ values are changed, in particular for large $r_+$
and $r_-$. Of course, this numerical results are just taken for the
low lying modes and the number of interaction are taken up to 50
times i.e. $\ell=50$, but it strongly indicate the highly damped
mode obtained by Stokes line methods is reliable at least $-r_-  \ge
r_+$.

\newpage

\end{document}